# Period of the d-Sequence Based Random Number Generator

## Suresh Babu Thippireddy and Sandeep Chalasani

**Abstract:** This paper presents an expression to compute the exact period of a recursive random number generator based on d-sequences. Using the multi-recursive version of this generator we can produce large number of pseudorandom sequences.

## Introduction

Random number generators (RNGs) are of utmost importance in many computer science applications [4-9]. One would like the period of the random sequence to be as large as possible, and also want it to have excellent autocorrelation properties [9].

Parakh has studied some properties of a recursive random number generator based on d-sequences [1,6] with the objective of having long periods. In this paper, we examine such RNGs afresh and provide some new results. For a background of d-sequences, see [2-4 and 9] and for approaches to the study of randomness in sequences, see [10]. Binary d-sequences may be easily generated using the formula [4]:

$$a(i) = 2^i \bmod p \bmod 2$$

where p is a prime.

## Recursive d-Sequence Generator

The recursive d-sequence based RNG that we use is given below:

$$(S^i \bmod p_{11} + S^i \bmod p_{12} + \ldots S^i \bmod p_{1n})^{k[i]} \bmod p_{21} \bmod 2$$
$$\oplus \quad (S^i \bmod p_{11} + S^i \bmod p_{12} + \ldots S^i \bmod p_{1n})^{k[i]} \bmod p_{22} \bmod 2 \ldots$$
$$\oplus \quad (S^i \bmod p_{11} + S^i \bmod p_{12} + \ldots S^i \bmod p_{1n})^{k[i]} \bmod p_{2m} \bmod 2 \qquad (1)$$

This is different from the one used by Parakh in as much as that "k" is replaced by "k[i]" which has a bearing on the calculation of the period of the random sequence.

The sequence may be generated by the algorithm given below:

**Algorithm:**

1) Select the seed element S to be a primitive element of all the primes $(p_{11}, p_{12} \ldots p_{1n})$.

2) Calculate the period of $(S^i \bmod p_{11} + S^i \bmod p_{12} + \ldots + S^i \bmod p_{1n})$, let us assume it to be '$t$'. Find the second level seeds $S_1, S_2, \ldots S_t$, using the above formula

3) Select the second level primes as $(p_{21}, p_{22}, p_{23}, \ldots, p_{2m})$

4) Calculate the period for each of the second level seeds with each second level primes i.e., from $i=1$ to t and store it in *k[i]*.

5) for $i=1$ to *t*

   Generate a sequence with a period *k[i]*, using the formula

   $(S^i \bmod p_{11} + S^i \bmod p_{12} + \ldots\ldots + S^i \bmod p_{1n})^{k[i]} \bmod p_{21} \bmod 2$ ⊕
   $(S^i \bmod p_{11} + S^i \bmod p_{12} + \ldots\ldots + S^i \bmod p_{1n})^{k[i]} \bmod p_{22} \bmod 2$ ⊕ …
   ⊕ $(S^i \bmod p_{11} + S^i \bmod p_{12} + \ldots\ldots + S^i \bmod p_{1n})^{k[i]} \bmod p_{2m} \bmod 2$.

6) We then would be able to generate a sequence of period 'L'. where

   $L = \sum k[i]$ ,where i=1 to t

**Calculation of the period at first level:**

1) Calculate the exact period all of the primes $p_{11}, p_{12}, \ldots . p_{1n}$ with the seed S. let the periods be $r_1, r_2, \ldots . r_n$.

2) Find the LCM of all the above periods and then store it in 't'.

   $t = LCM\ (r_1, r_2 \ldots r_n)$.

**Calculation of periods at the second level and length of the sequence:**

1) Calculate the exact period of each second level primes with the second level seed and find the LCM of the above period ,store it in k[i].
   i.e.,

   $k[i] = LCM(period(p_{21}, S_i), period(p_{22}, S_i), \ldots., period(p_{2m}, S_i))$.

   Then the period of the sequence 'L' is found by the Summation of the above periods.

   $L = \sum k[i]$, where i=1 to t

In the examples below we confine ourselves to simple cases dealing with just a few primes. It can be readily extended to cases where the number of primes is many and where the size of the primes is large.

**Examples:**

1) Let us assume that  $S=2$, $p_{11}=3$, $p_{12}=5$, $p_{21}=7$, $p_{22}=11$

We get the second level seeds as $S_1=4$ , $S_2=5$ , $S_3=5$, $S_4=2$.

Then for the above seeds with the second level primes we get periods as,

k[1]=15, k[2]=30, k[3]=30, k[4]=30

Then calculate the summation of above periods to get the length of the sequence which is same as the period of the sequence. Here we get it as L=105.

2) If S=2 , $P_{11}$ = 23, $P_{12}$=29, $P_{21}$=47, $P_{22}$=53

We have 308 number of second level seeds, then for the above seeds we get 308 periods. By summation of the above periods we will get the period of the entire sequence as 253253

## Multi-recursive random number generator

An expression for a multi-recursive RNG is given below:

$$((S^i \bmod p_{11}+S^i \bmod p_{12}+\ldots S^i \bmod p_{1n})^{k[i]} \bmod p_{21} \ +$$
$$(S^i \bmod p_{11}+S^i \bmod p_{12}+\ldots S^i \bmod p_{1n})^{k[i]} \bmod p_{22} \ldots +$$
$$(S^i \bmod p_{11}+S^i \bmod p_{12}+\ldots S^i \bmod p_{1n})^{k[i]} \bmod {}_{2m})^{v[i]} \bmod p_3 \bmod 2 \qquad (2)$$

The algorithm for this same as the above with an additional step for calculating the period of the third level seeds with the third level prime and storing it in an array v[i] , here i=1 to L. The operation $\oplus$ is replaced by the '+' operation.

The length of the sequence obtained here is $\sum v[i]$, where i=1 to L, but the period remains same as the period of the recursive d-sequence.

The RNG obtained using equation (2) may be further generalized by adding further nested layers of the modulus operation.

**Examples**:

1) Let $p_{11}$= 11, $p_{12}$=13 ,$p_{21}$=5 , $p_{22}$=7

   for $p_3$=7
   we get the period= 367   and the length of the sequence =1059

   for $p_3$=13
   we get the period= 367  and the length of the sequence =2066

   for $p_3$=19
   we get the period=367 and the length of the sequence =3373

   for $p_3$=31
   we get the period=367 and the length of the sequence =4398

2) Let $p_{11}$=17, $p_{12}$=19 ,$p_{21}$=7, $p_{22}$=11

   for $p_3$= 5
   we get the period= 855 and the length of the sequence =2531

for p$_3$=11
we get the period=855 and the length of the sequence =5323

for p$_3$=19
we get the period=885 and the length of the sequence =8995

for p$_3$=31
we get the period=885 and the length of the sequence =10296

3) Let p$_{11}$= 3, p$_{12}$=5, p$_{21}$=7, p$_{22}$=11

for p$_3$= 7
we get the period=105 and the length of the sequence =412

for p$_3$=11
we get the period=105 and the length of the sequence =729

for p$_3$=17
we get the period=105 and the length of the sequence =1332

for p$_3$=23
we get the period= 105 and the length of the sequence = 1672

The measure of randomness of a discrete sequence x is *R(x)* given by the expression below:

$$R(x) = 1 - \frac{\sum_{k=1}^{n}\left|C(k)\right|}{n}$$

where *C(k)* is the autocorrelation value for *k* and *n* is the period of sequence. The value of the autocorrelation is defined as in the equation below:

$$C(k) = \frac{1}{n}\sum_{j=0}^{n} a_j a_{j+k}$$

Here are some Tables and graphs for recursive random number generator with seed =2 and the primes and period labeled.

**Table 1:** Randomness measure of different recursive sequences

| Seed | $P_{11}$ | $P_{12}$ | $P_{21}$ | $P_{22}$ | Period | Randomness |
|------|------|------|------|------|--------|------------|
| 2 | 3 | 5 | 19 | 17 | 396 | 0.953015 |
| 2 | 3 | 5 | 23 | 29 | 770 | 0.976219 |
| 2 | 5 | 7 | 19 | 17 | 1056 | 0.972376 |
| 2 | 11 | 13 | 29 | 31 | 9625 | 0.990949 |
| 2 | 3 | 5 | 7 | 11 | 105 | 0.872200 |
| 2 | 23 | 29 | 7 | 11 | 4112 | 0.984785 |

Here in each table the initial primes are kept same and the second level primes are changed. The seed value is Seed=2.

**Table 2:** Varying period by changing second-level primes

| $P_{11}$ | $P_{12}$ | $P_{21}$ | $P_{22}$ | Period |
|------|------|------|------|--------|
| 3 | 5 | 7 | 11 | 105 |
| 3 | 5 | 19 | 17 | 396 |
| 3 | 5 | 23 | 29 | 770 |
| 3 | 5 | 31 | 37 | 342 |
| 3 | 5 | 41 | 43 | 1050 |
| 3 | 3 | 47 | 53 | 4186 |
| 3 | 5 | 71 | 73 | 1350 |
| 3 | 5 | 79 | 83 | 11193 |
| 3 | 5 | 101 | 103 | 12750 |
| 3 | 5 | 107 | 109 | 8586 |
| 3 | 5 | 113 | 127 | 714 |
| 3 | 5 | 131 | 137 | 24310 |
| 3 | 5 | 139 | 149 | 20424 |
| 3 | 5 | 163 | 167 | 29133 |

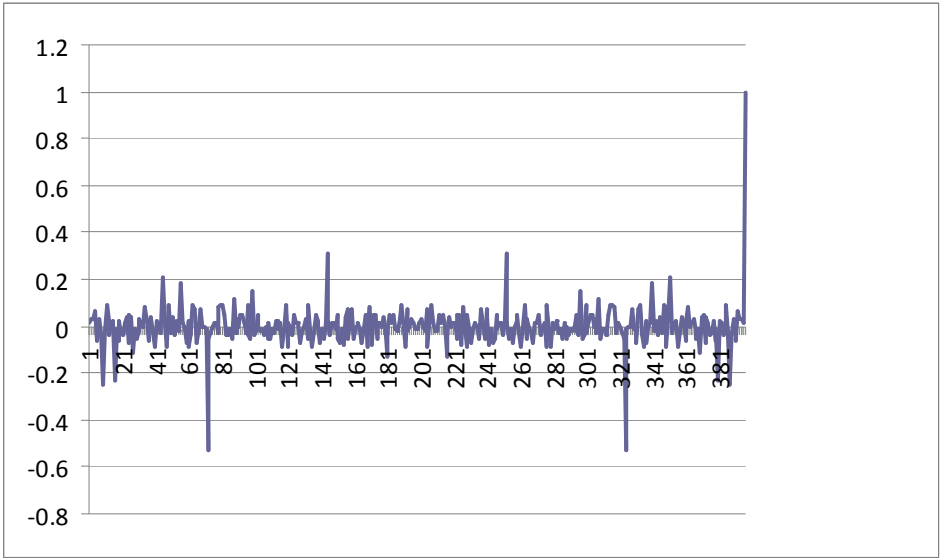

**Figure 1:** Autocorrelation graph for primes 3-5-17-19

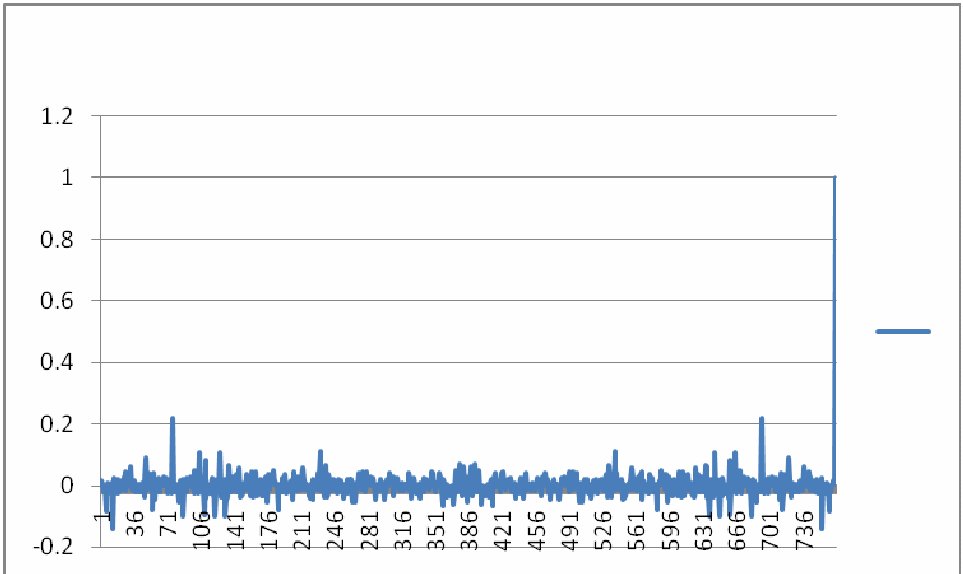

**Figure 2**: Autocorrelation graph for 3-5-23-29

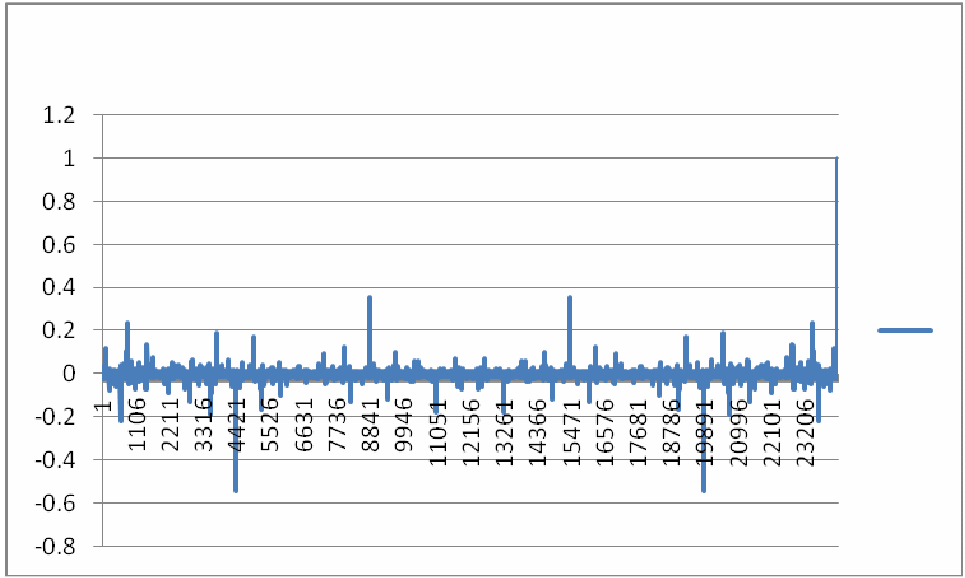

**Figure 3**: Autocorrelation graph for 3-5-131-137

**Table 3:** Varying period by changing second-level primes

| P_{11} | P_{12} | P_{21} | P_{22} | Period |
|--------|--------|--------|--------|--------|
| 23 | 29 | 7 | 11 | 4112 |
| 23 | 29 | 19 | 17 | 17706 |
| 23 | 29 | 31 | 37 | 21093 |
| 23 | 29 | 41 | 43 | 84595 |
| 23 | 29 | 47 | 53 | 253253 |

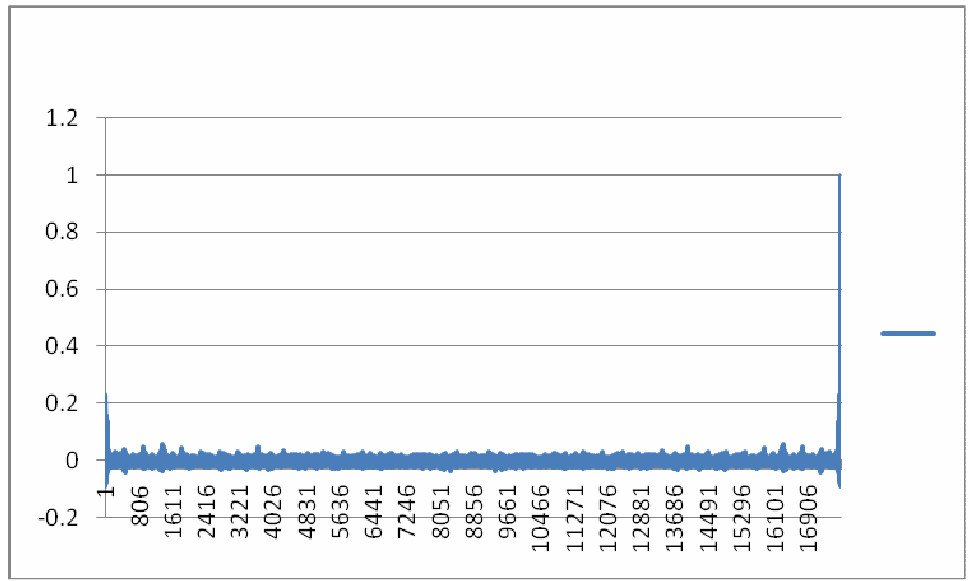

**Figure 4**: Autocorrelation graph for Primes 23-29-19-17

From the above tables we can say that by keeping the second level primes $p_{21}$, $p_{22}$ constant and increasing the initial primes $p_{11}$, $p_{12}$ the period can be increased

## Conclusion

We have provided an expression for computing the period of the d-sequence based recursive RNG, which is somewhat different from that of Parakh. We have presented results summarizing the properties of such an RNG.

The multi-recursive RNG described in this paper can provide us a flexible approach to generating random sequences with varying and large periods. Extensions of this work will be to investigate recursive versions of the cubic transformation [8] and the application of one-way transformations to each step of the recursive mapping [7].